\documentclass[debug]{rmaa}

\usepackage{paralist}

\usepackage{psfrag,color}

\usepackage[latin1]{inputenc}
\usepackage{natbib}

\usepackage{float}

\title{NGC 2440 : A morpho-kinematical model} 

\author{
  P.J.A. Lago,\altaffilmark{1}
  R.D.D. Costa,\altaffilmark{1}
}
\altaffiltext{1}{IAG, USP, Brazil} 

\shortauthor{Lago \& Costa}
\shorttitle{NGC 2440 : A morpho-kinematical model}

\fulladdresses{
\item Lago, P.J.A., Costa, R.D.D.: Instituto da Astronomia, Geof\'isica e Ci\^encias Atmosf\'ericas, Universidade de S\~ao Paulo, Rua do Mat\~ao 1226, Cidade Universit\'aria. 05508-090 - S\~ao Paulo/SP, Brazil (paulolago@usp.br; roberto.costa@iag.usp.br).}

\listofauthors{Lago, P.J.A. \& Costa, R.D.D.}
\indexauthor{Lago, P.J.A.}
\indexauthor{Costa, R.D.D.}

\abstract{

This work describes the modelling of the 3D structure and position-velocity (P-V) diagrams of NGC 2440, a well known planetary nebula, aiming to describe the morphology of this object, specially its core. We have used high resolution spectra and P-V diagrams to reproduce the 3D structure of the nebula using SHAPE, a  software that allows 3D modelling. HST high angular resolution images were used as reference to the model. 
 
The observational data point to a segmented core, and the simulations confirm this assumption; the best model for the nebula is a torus segmented in three pieces. The simulated P-V diagrams agree with the observations. We suggest that the torus was torn apart by interaction with the surrounding medium, either as winds or the radiation field.

For the two bipolar lobes, orientation angles to the plane of the sky of $27\pm 5$ and $-5\pm3$ degrees, respectively for the bipolar components with PA of 85 and 35 degrees, were derived. No additional bipolar lobes were required to model the observed features of  NGC 2440. A distance of $1.8 \pm 0.5 kpc$ was derived for the nebula using our velocity field for the toroidal structure. These results are the first derived for NGC 2440 from modelling in a 3D environment.

}
\resumen{}

\addkeyword{planetary nebulae: individual (NGC 2440)}
\addkeyword{techniques: spectroscopic}
\addkeyword{methods: numerical}
\addkeyword{ISM: kinematics and dynamics}

\begin{document}
\maketitle

\section{Introduction}
\label{sec:intro}

The study of planetary nebulae morphology is the key to understand their formation process; using specific information for a given object, it is possible to apply a morpho-kinematical model, taking into account projecting effects, and derive physical properties such as kinematical age, that can be derived from the velocity field and the observed geometry. Morphological classification schemes for PNe started very early, the first one was developed using a sample of 78 PNe by \citet{1918PLicO..13...55C}. The first schemes were based only on nebular shapes, later on other schemes were proposed, based on the central star position in the HR diagram, such as that by \citet{1993A&A...279..521S}. Present works consider also projection effects to better describe the morphological features of the objects.

Aiming to improve the understanding of projection effects in morphological classification, \citet{1998ApJS..117..341Z} have developed a modelling method that produces automatically, based on a set of parameters, possible scenarios for elliptical PNe. Similarly, \citet{2012ApJ...760..115C} have shown that multi-polar PNe are more common than previously supposed, based on a set of simulations, varying the angles between three identical bipolar components. All diversity of morphologies were generated by projection effects and relative position of each bipolar component, so their conclusion is that the quantity of multi-polar PNe was underestimated.

The origin of multi-polar structures in planetary nebulae is, in general, associated to binarity of the central star, as described by \citet{1994ApJ...421..219S}. In this scenario the object is constituted by a close binary system whose members share a common envelope, and the mass ejections occur preferentially in the orbital plane. In view of this effect, densities in the orbital plane and at the poles are different, and when the hot core is exposed a fast and diluted wind, in interaction with the medium, shapes the nebula.

The multi-polar structure can be explained by multiple ejection episodes, with different axes of symmetry; in planetary nebulae this kind of ejection is described by a mechanism known as bipolar rotating episodic jets (BRETs) that consist in a collimated flow that could produce shock-excited regions \citep{2013A&A...557A.121G}. These regions are associated to the front of the outflow, which in turn are episodic; observations of these phenomena suggest that exists precession or rotation in the jet \citep{1995ApJ...455L..63L}.

There is however an alternative to explain a multi-polar structure without BRETs mechanism. \citet{2013MNRAS.436..470S} performed a numerical simulation aiming to create a multi-polar structure in a PN without BRETs; their work shown the possibility to form this kind of structure by the interaction of expanding shells, each one with different densities, with the outer shell containing a filamentary distribution of mass. Bipolar and starfish PNe are well reproduced using this technique, but it is important to point out that these two scenarios can coexist, the last one is more suitable to explain not well collimated structures, completing the BRETs hypothesis.

NGC 2440, a multi-polar object with at least two bipolar components, is a well known planetary nebula. \citet{1998ApJ...493..803L} described in great detail its main structure and some of the substructures. There are clearly two bipolar structures, the largest one has a position angle (PA) of 85 degrees, and the smaller has a position angle of 35 degrees, so that they are tilted 50 degrees apart. A third bipolar component was also suggested but, due the lack of observational evidence, its existence is speculative; this younger bipolar component would have a PA of 60 degrees, an inclination based on the position of two central knots, and would be tilted with respect to the other bipolar components.

An important feature of NGC 2440 is the molecular emission, in particular for the bipolar component with PA of 35 degrees. This emission, measured in CO J=3-2, coincides with the optical component, and the maximum line fluxes are associated to the knots at the extremity of the bipolar component, as described by \citet{2008ApJ...673..264W}, who measured molecular outflows with different velocities. In the core region, the presence of fast CO components points to the destruction of a neutral and dense torus by the outflow at the moment of the lobes formation.

The central region is complex and rich in micro-structures. An image analysis displays the core surrounded by a toroidal configuration, associated to many radial filaments \citep{1998ApJ...493..803L}, a description that agrees well with the results of \citet{1991ApJ...377..210R}, that suggest the presence of dense and small clumps surrounded by an optically thinner medium.

The SHAPE modelling tool \citep{2010arXiv1003.2012S} allows to reconstruct the 3D structure of a planetary nebula in great detail, taking into account projection effects, using high angular resolution images such as those from the Hubble Space Telescope (HST), and high resolution spectra. This kind of modelling has shown results in  good agreement with observational data, revealing important morpho-kinematic features for many objects such as those described by \citet{2012ApJ...761..172G}, \citet{2014Msngr.158...26S} and \citet{2015MNRAS.446.1931R}, among others.

The goal of the present work is to present the first model based on SHAPE for NGC 2440. This model aims the reproduction of the overall 3D structure of the nebula, improving the current models for this object.
In section 2 the observational data used in the work are described; in section 3 the setup of the modeling tool for the simulations is discussed; in section 4 and 5 the simulation results are compared to the observational data and the morphology of the nebula is discussed, and finally, in section 6 our conclusions are drawn. 

\section{Observational data}

The San Pedro M\'artir (SPM) Kinematic Catalogue of Planetary Nebulae is a large and specialized survey of PNe, constituted of a database of high resolution long slit spectra of galactic and extragalactic PNe. The instrumentation used to get the data were the Anglo-Australian 3.9m and San Pedro M\'artir 2.1m telescopes. Details of the survey are described by \citet{2012IAUS..283...63L} and the data are available online\footnote[1]{\textit{http://kincatpn.astrosen.unam.mx/index.}}. This database is an excellent source in view of high uniformity of the data, besides its high instrumental sensibility to nebulae.

\begin{figure}[htb]
\begin{center}
\includegraphics[scale=0.25]{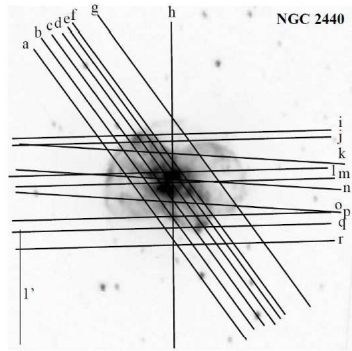}
\caption{Slit positions from SPM survey to NGC 2440.}
\end{center}
\end{figure}

Aiming at the construction of the P-V diagrams for NGC 2440, each frame was corrected by the systemic velocity, adopting the value of $-65\,km/s$ from \citet{1998ApJ...493..803L}. Figure 1 shows the slit positions for NGC 2440 available in SPM catalogue.

\section{Modeling the 3D structure}

The 3D morpho-kinematic model aims to describe the structure of NGC 2440, taking into account the projection effects, and therefore improving previous results such as those by \citet{1998ApJ...493..803L}, verifying the orientation angles, as well as describing in more detail the central region.

The SHAPE modelling tool allows to build a model, based in a mesh model inside a 3D environment, in order to analyse the morphology and kinematics of PNe taking into account projection effects. The structure is build manually, in an approach similar that is used in graphical computation softwares, as described by \citep{2010arXiv1003.2012S}. The construction of the structure is based on the interpretation of observational data, that are high resolution long slit spectra, and high angular resolution images. It has to be noted that SHAPE does not calculate the radiative transfer equation for the problem. The geometry of NGC 2440 was established based on [NII] HST images. Figure 2 displays the HST image and the mesh model used in this work for the sake of comparison.

The evaluation of the model is made by comparing the observational data directly with the SHAPE output, that are two rendered images, one simulating the photometric image of nebula and other simulating the P-V diagram derived from a long slit spectrum. Our work was aimed to reproduce the P-V diagrams from [NII] lines, the spectral range for which we have the best sampling. SHAPE also provides tasks that simulate the observational conditions such as seeing and instrumental setup.

SHAPE modelling requires a mesh model and a set of physical parameters such as  velocity fields and densities distributions, derived from the observational data. In general, the larger is the database, the better is the model due to the increasing in the number of independent constraints.

An important aspect of the present model is that it considers the nebula as formed by symmetrical structures. The reproduction of small asymmetries like cavities, filaments and small knots that can be seen in optical images is out of the scope of this paper. The high angular resolution images required for the modelling of NGC 2440 were selected from HST Wide Field Planetary Camera (WFPC2), available in different filters \footnote[3]{\textit{http://mast.stsci.edu/portal/Mashup/Clients/Mast/Portal.html}}. It is important to emphasize that the HST images, in particular those acquired using filter F658N, were used only as reference for the construction of the mesh model. The reproduction of the photometric images in all their details, including the complex filamentary structures existing in this nebula, is not the goal of the present work, since our model mainly aims to describe the P-V diagrams.

The two bipolar components were considered symmetric and each position angle was adjusted in order to reproduce the reference image. Besides these two main components, five other features are also reproduced: two knots, clearly visible at the edges of the smaller bipolar component at PA 35 degrees, as already shown by \citet{1998ApJ...493..803L}, and the central region subdivided in three, non aligned, torus pieces. 

\begin{figure}[H]
\begin{center}
\includegraphics[scale=0.25]{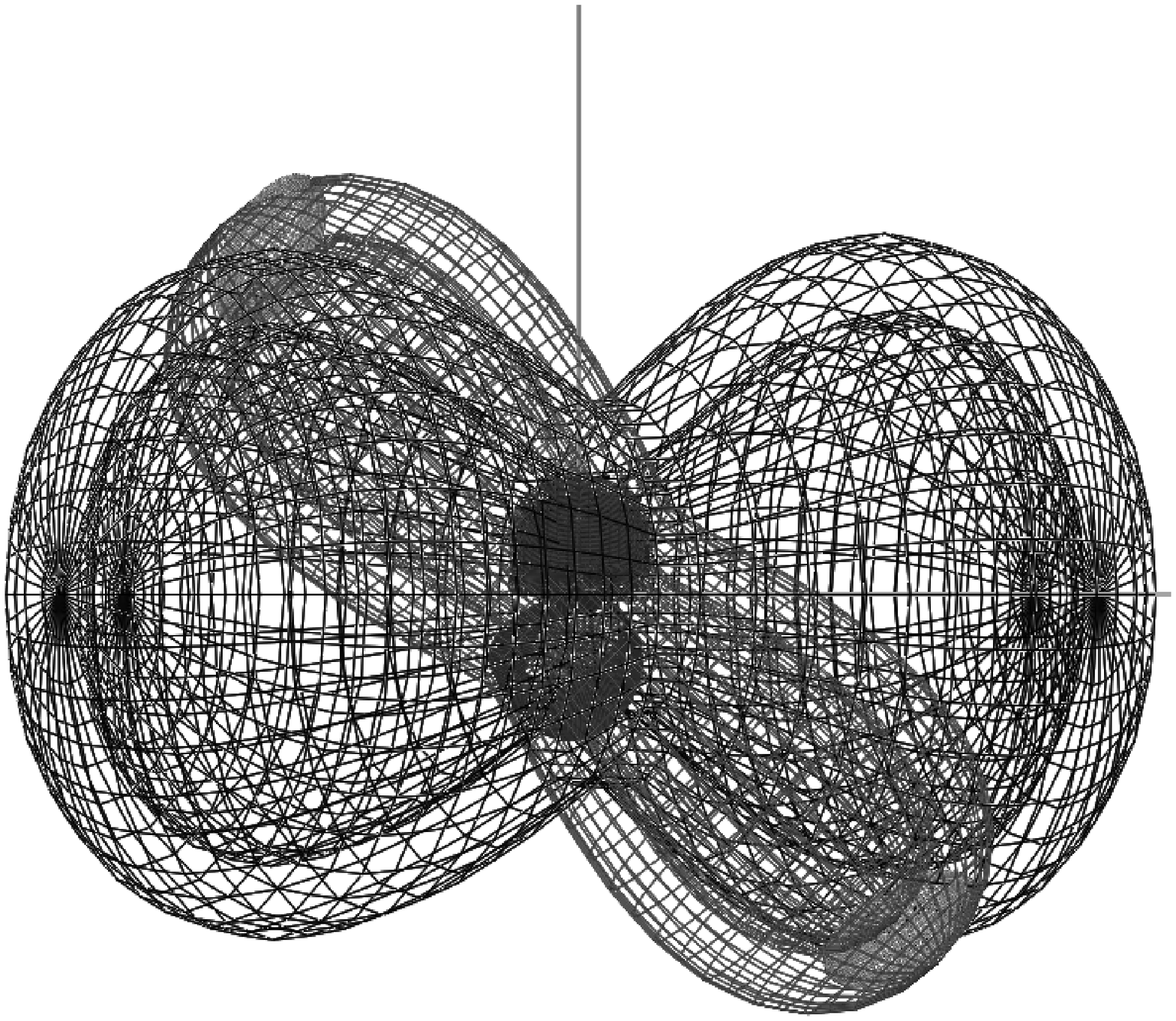}
\includegraphics[scale=0.30]{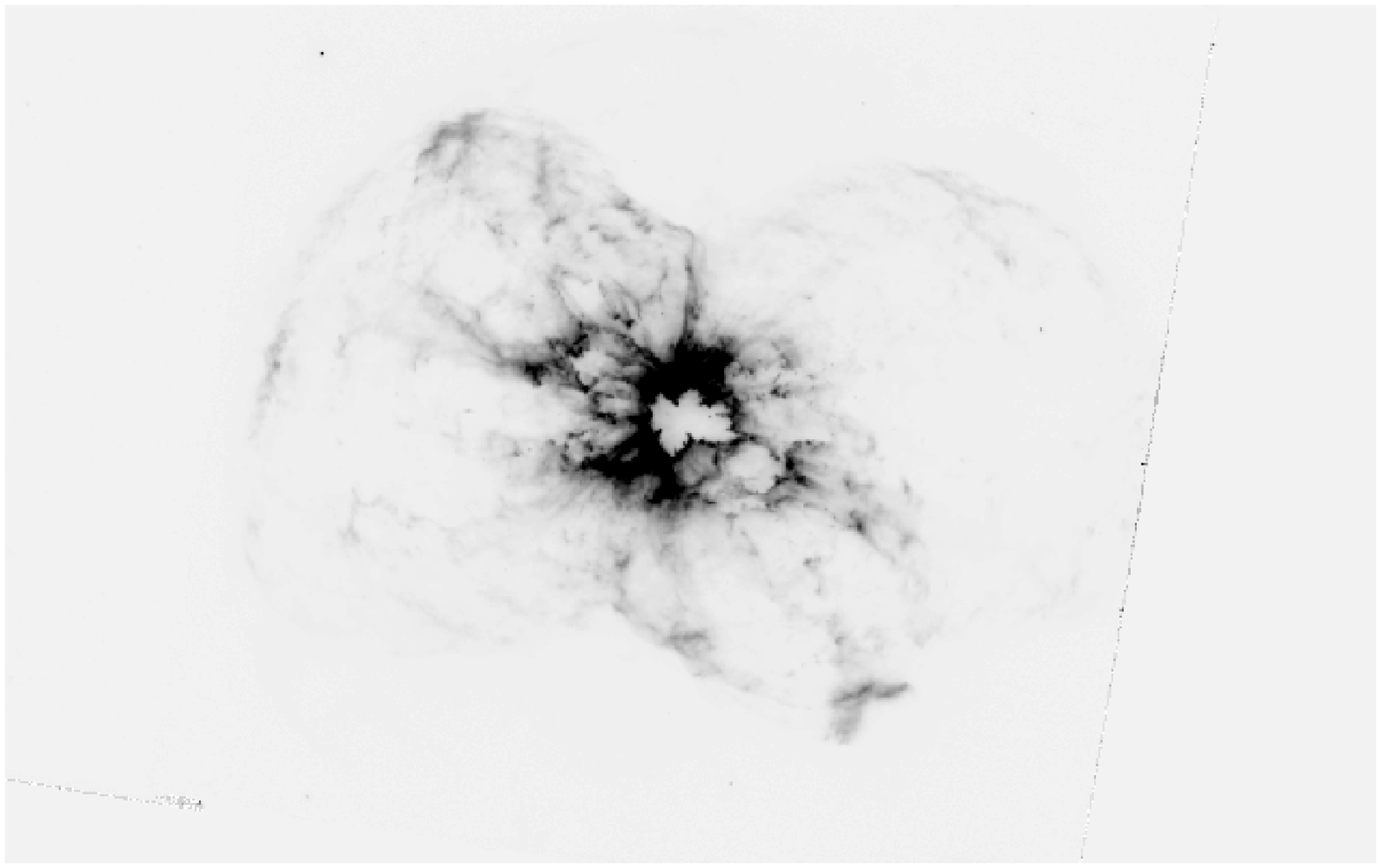}
\caption{Mesh model to NGC 2440 built with SHAPE (left) and reference HST image (right).}
\end{center}
\end{figure}

The choice of the velocity field is an important step of the modelling process since the shape of the P-V diagram will be affected by this parameter, and it should reproduce the observed data, in combination with the geometry, taking also into account the projection effect.

The adopted velocity fields were linear, increasing with the distance to the core. Table 1 shows the list of parameters of each velocity field used, for the different components, and the expression for the velocity field is shown in equation 1, where $r$ is the distance to the center, $r_{o}$ is the scale factor, in general the size of structure, and $k$ is a constant. It should be emphasized that for the knots $r_{o}$ is the distance to the center, not the size of the structure.

We chose a linear field, that implies in a ballistic outflow assumption,  based not only in its simplicity but also based on previous works in the field, such as those by \citet{2012MNRAS.423..925A}, \citet{2013MNRAS.434.1505H}, \citet{2015A&A...582A..60C} and \citet{2016MNRAS.457.3409A}, that use the same approach. \citet{2014Msngr.158...26S} use a non linear velocity field to describe part of their target object (the Homunculus nebula in Eta Carinae), but in this case the field is almost linear and aims to describe in great detail the deviations from linear fields. \citet{2014AN....335..378S} conclude that the velocity fields are almost never linear, but in their specific case the expansion law is a consequence of hydrodinamical codes, considering the time dependence of velocities. Therefore the present work adopts, as a first approximation, a linear velocity field, in agreement with those observed for the expansion law in bipolar PNe, as described by \citet{2004ASPC..313..148C} and \citet{2011MNRAS.416..715S}.

\begin{table}[htb] 
\centering 
\small 
\setlength{\arrayrulewidth}{1\arrayrulewidth} 
\setlength{\belowcaptionskip}{10pt} 
\caption{Model velocity field parameters.}
\begin{tabular}{|c|c|c|c|c|c|} 
\hline
Component & $k (km.s^{-1})$ & $r_{o}$ (arcsec)\\ 
\hline                               
lobe 85 & 184 & 34.0\\
lobe 35 & 160 & 30.1\\
knot N & 80 & 29.4\\
knot S & 84 & 30.8\\
torus piece I & 53 & 10.5\\ 
torus piece II & 44 & 7.8\\
torus piece III & 52 & 9.1\\ 
\hline       
\end{tabular}
\label{tab:Referencia_desejada}
\end{table}

Different density distributions were selected for each component in order to reproduce line contrasts in the P-V diagrams. As mentioned before, equation 1 below shows the expansion law.

\begin{equation}
v(r)=k\frac{r}{r_{o}}
\end{equation}

\section{Results}

Figures 3 to 8 show the kinematical profiles for different slit positions for [NII] line displayed in Fig. 1. In each figure the observed P-V diagram (left) is compared to the simulated one (right); the horizontal axis represents the heliocentric velocity. Since SHAPE produces a model in the reference frame of the PN, horizontal axis represents the expansion velocity corrected by the proper motion, such that both observational an simulated data are in the same reference frame.

Comparing the results of the simulations and the observed kinematical line profiles, it can be seen that the main structures of NGC 2440 were reproduced, in a good agreement with the observational data. Optimization and choice of the best model were made by comparing the results and the observational data.

\begin{figure}[H]
\begin{center}
\includegraphics[scale=0.4]{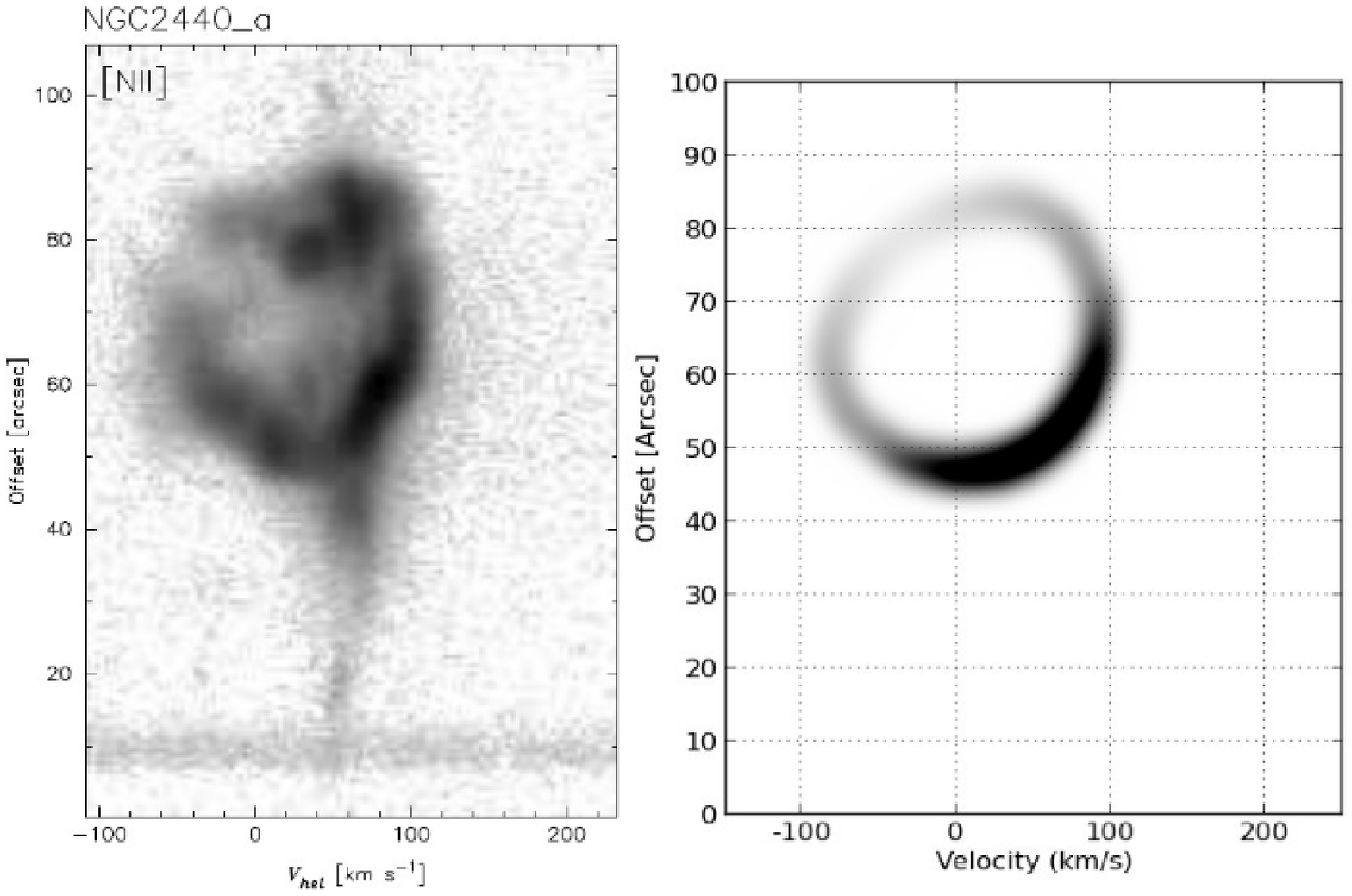}
\includegraphics[scale=0.4]{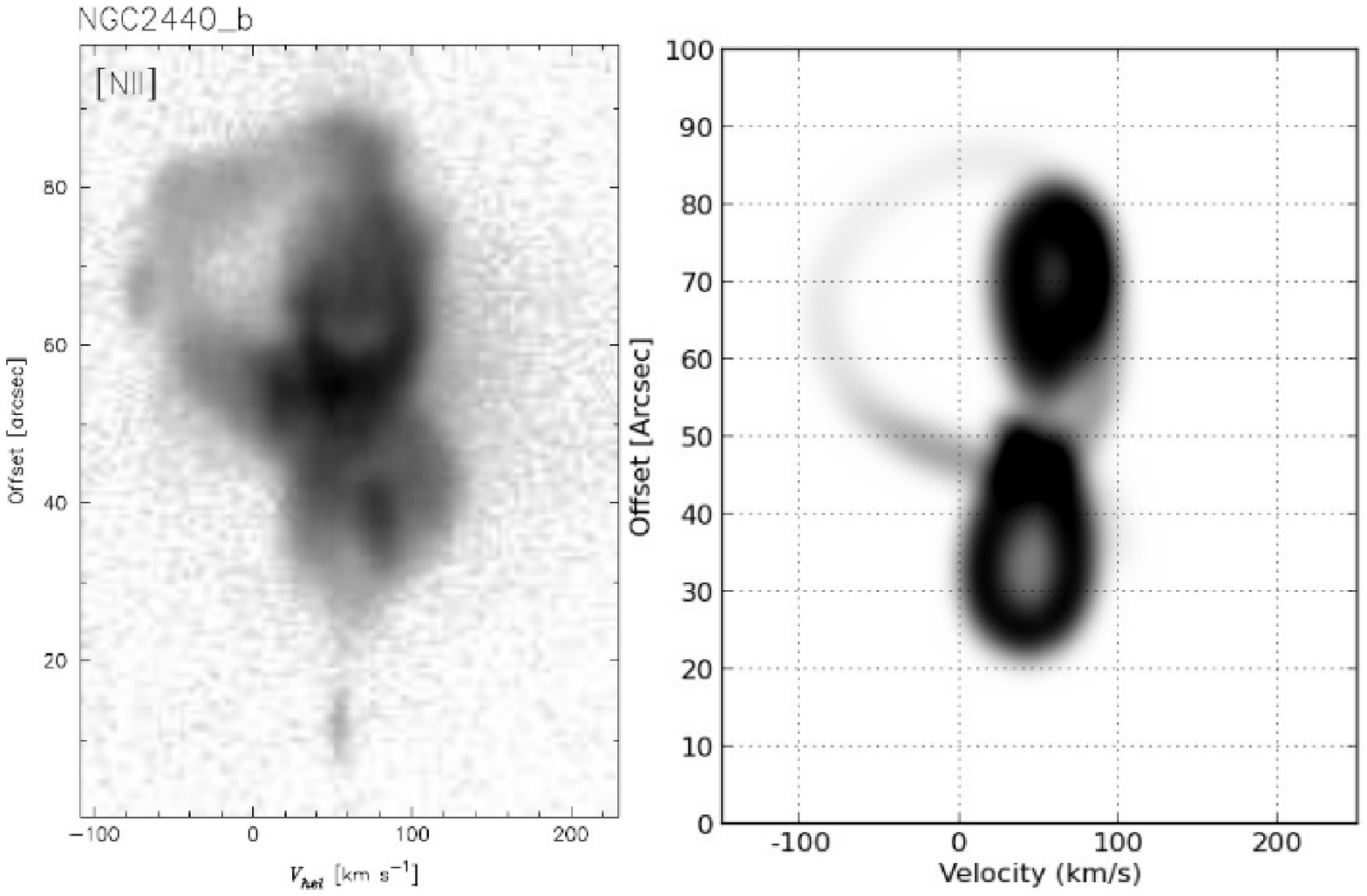}
\includegraphics[scale=0.4]{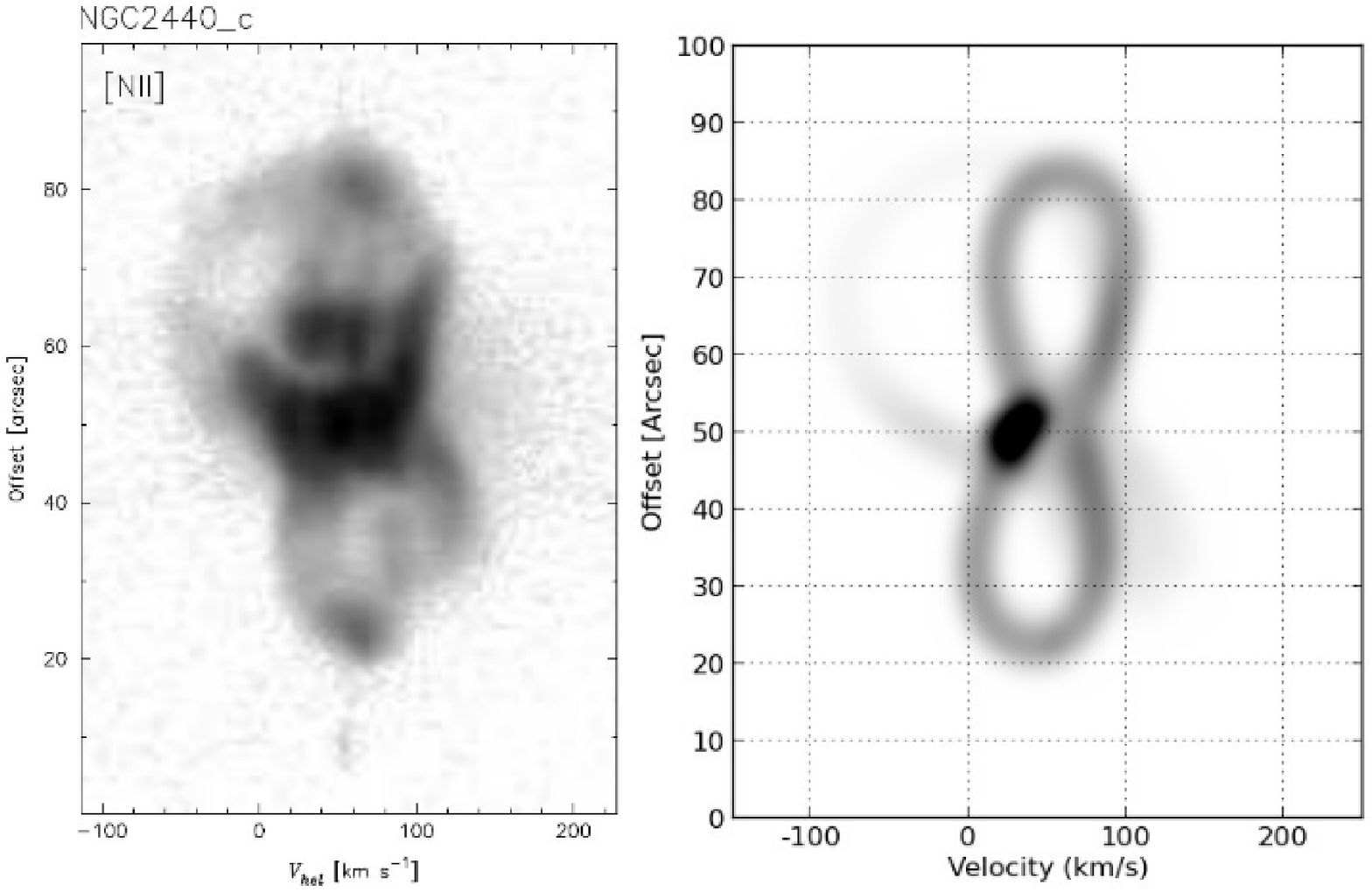}
\caption{P-V diagrams for slit positions a, b, c for NGC 2440. Observed profiles (from SPM survey) are in the left painel, and model outputs in the right painel. For the observed diagrams, values in the horizontal axis represent the heliocentric velocity.}
\end{center}
\end{figure}

\pagebreak

\begin{figure}[H]
\begin{center}
\includegraphics[scale=0.4]{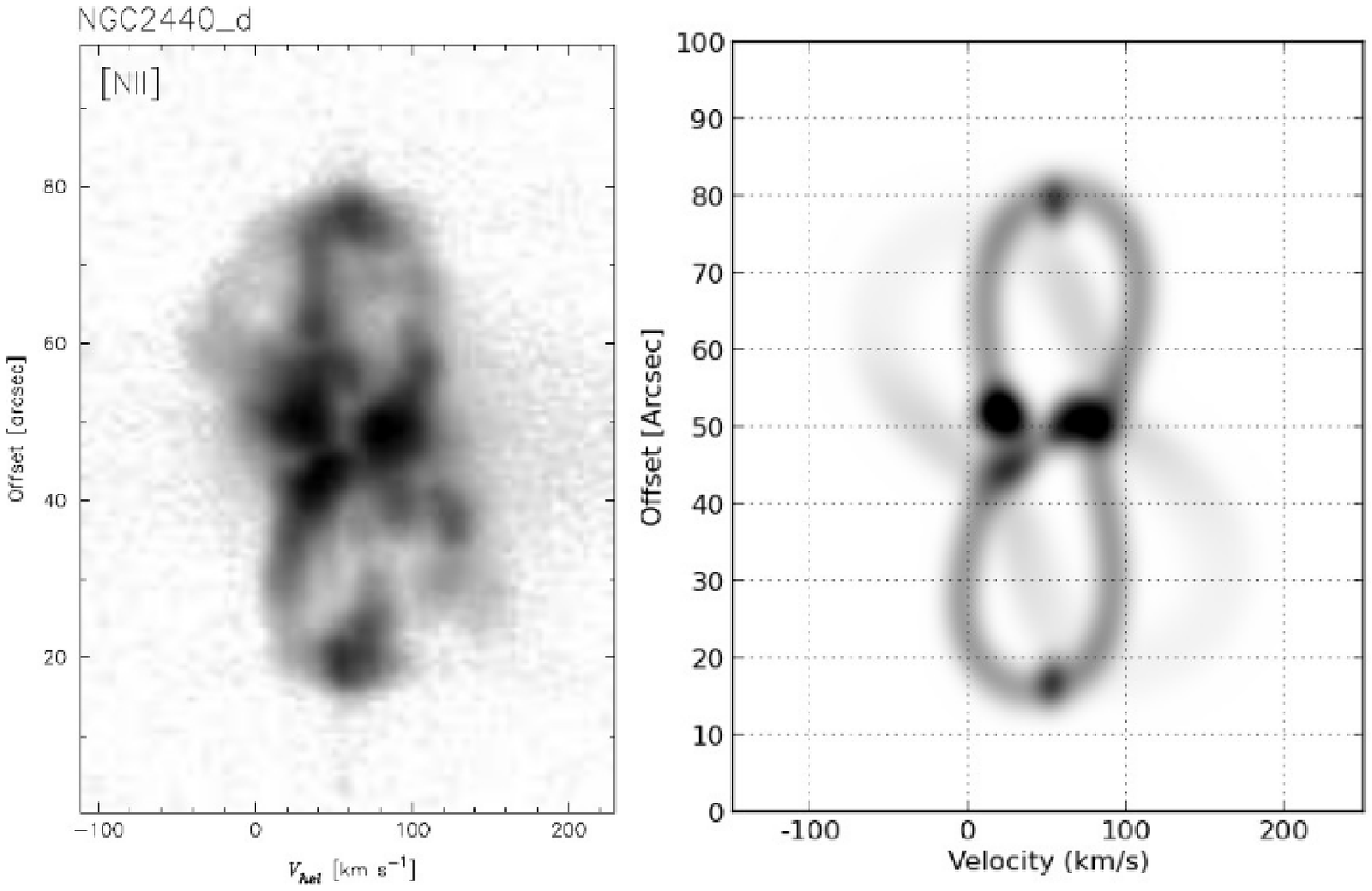}
\includegraphics[scale=0.4]{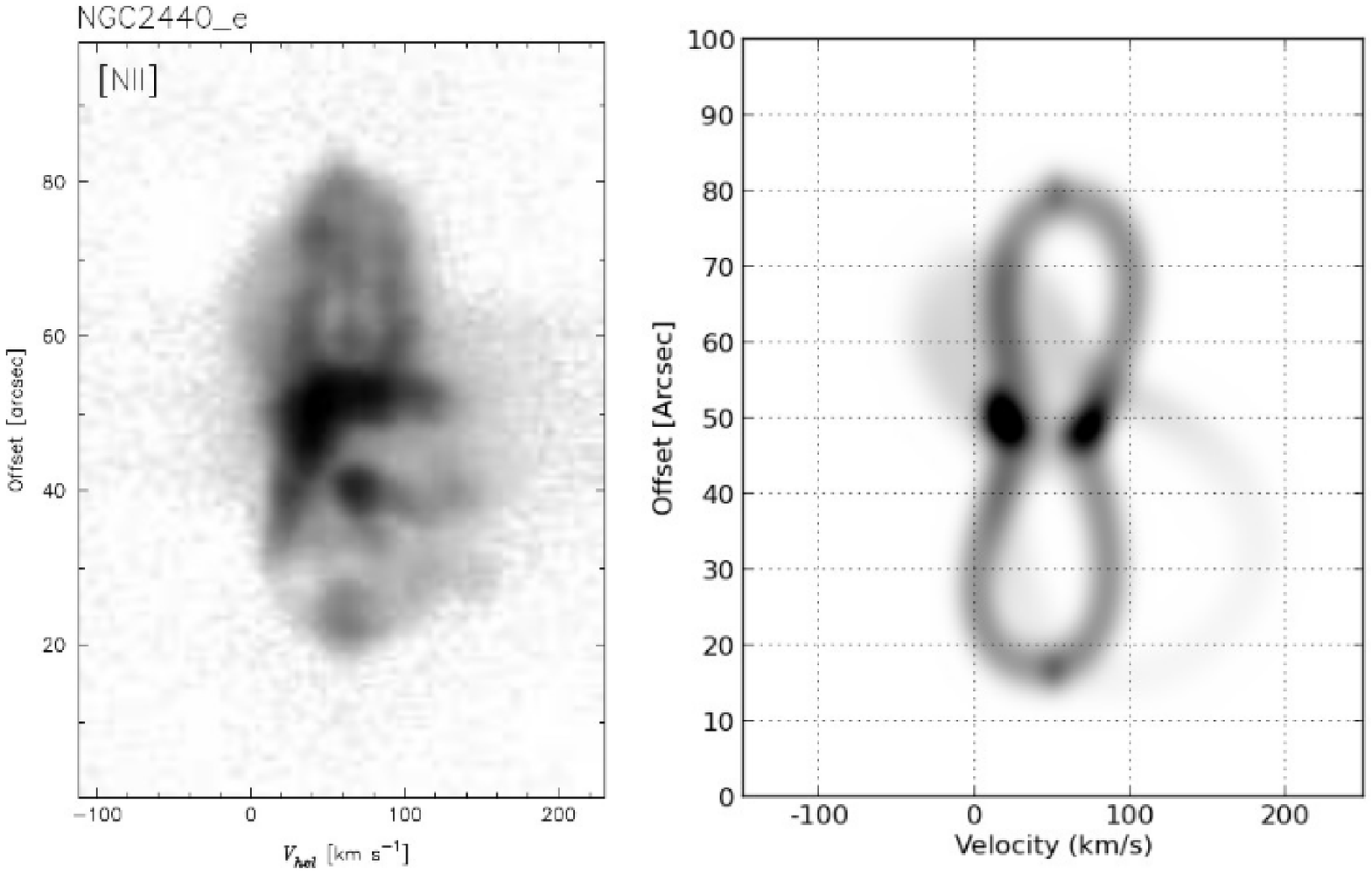}
\includegraphics[scale=0.4]{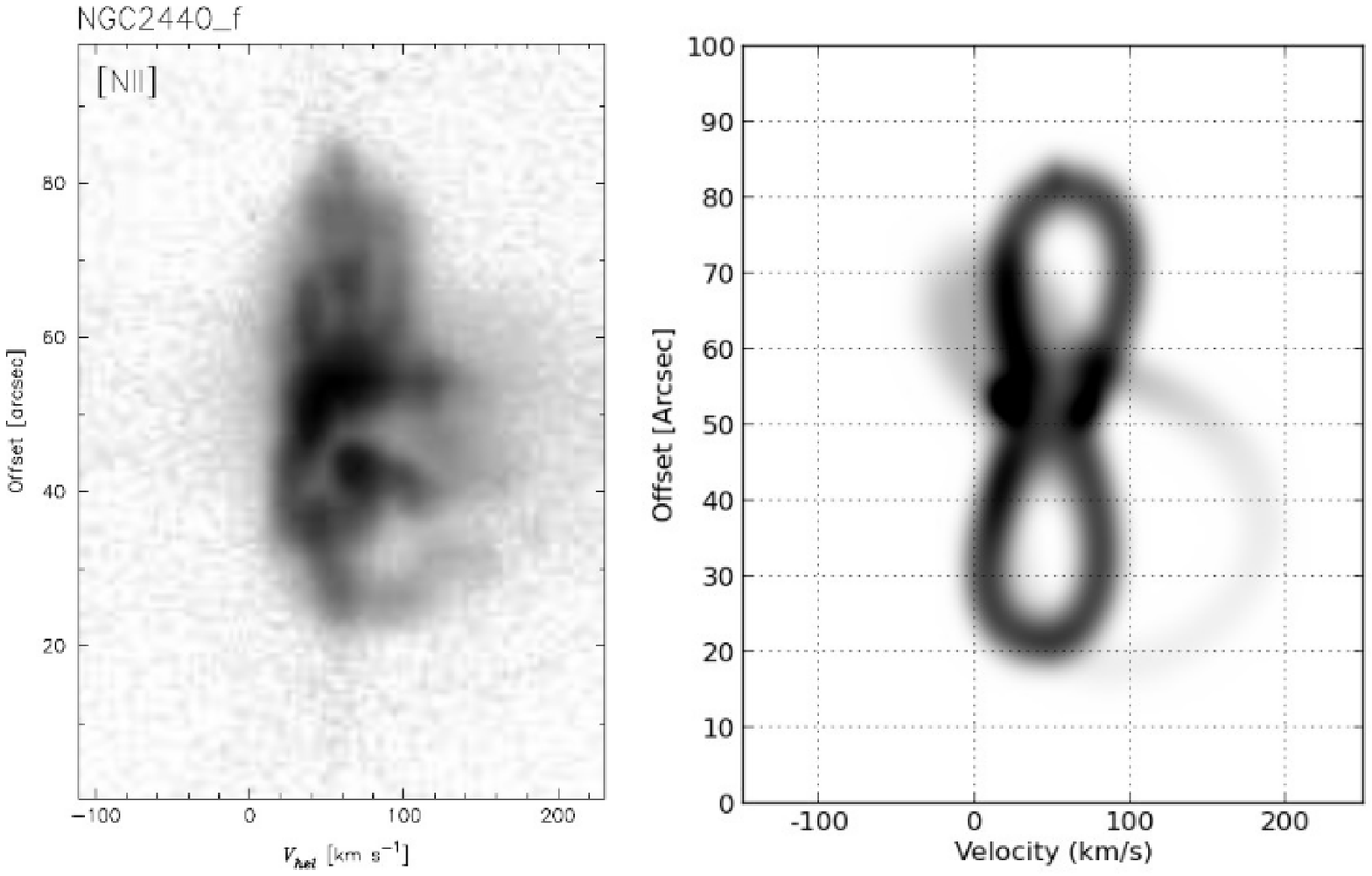}
\caption{Same of figure 3, for slit positions d, e and f.}
\end{center}
\end{figure}

\pagebreak

\begin{figure}[H]
\begin{center}
\includegraphics[scale=0.4]{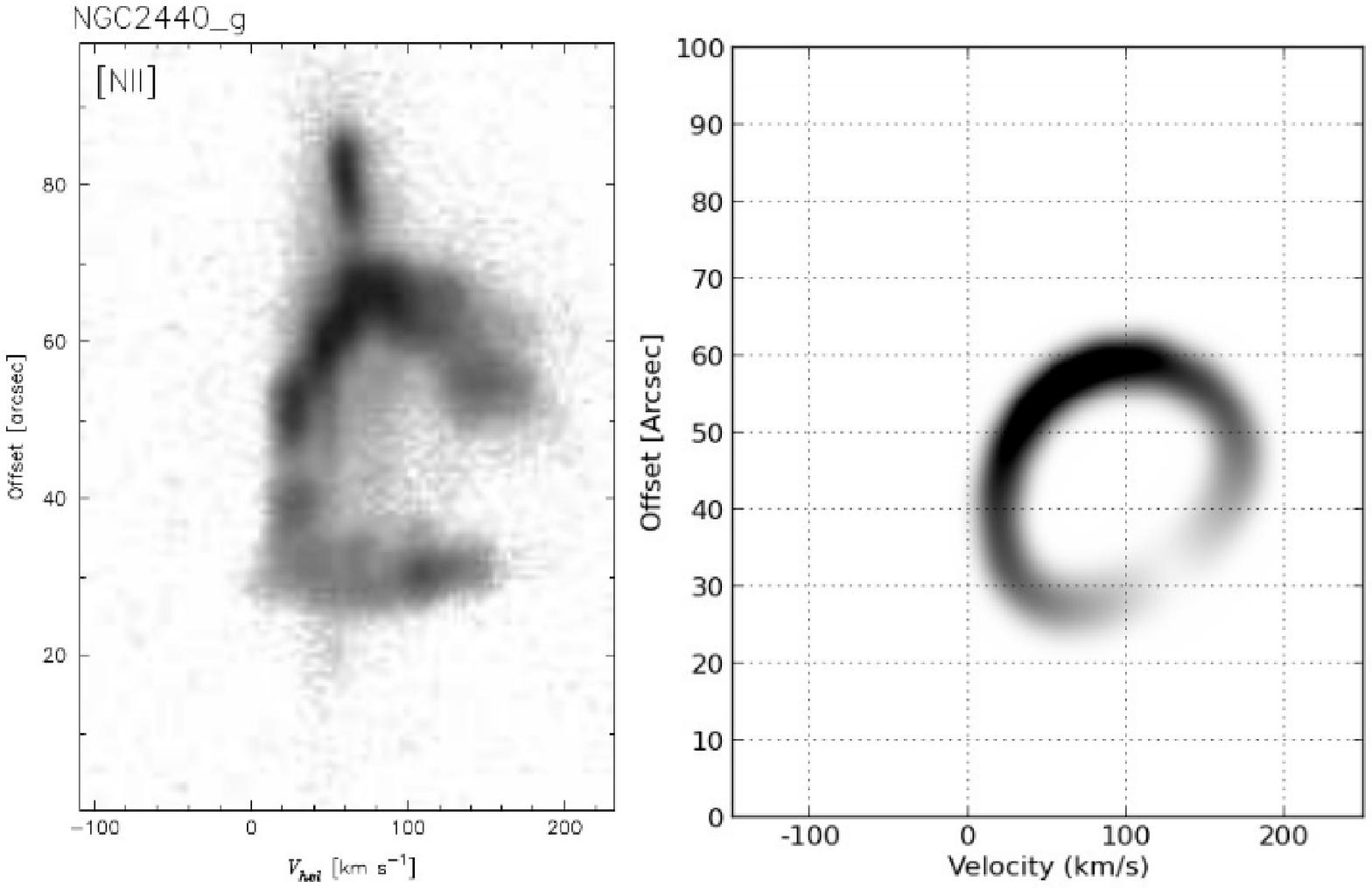}
\includegraphics[scale=0.4]{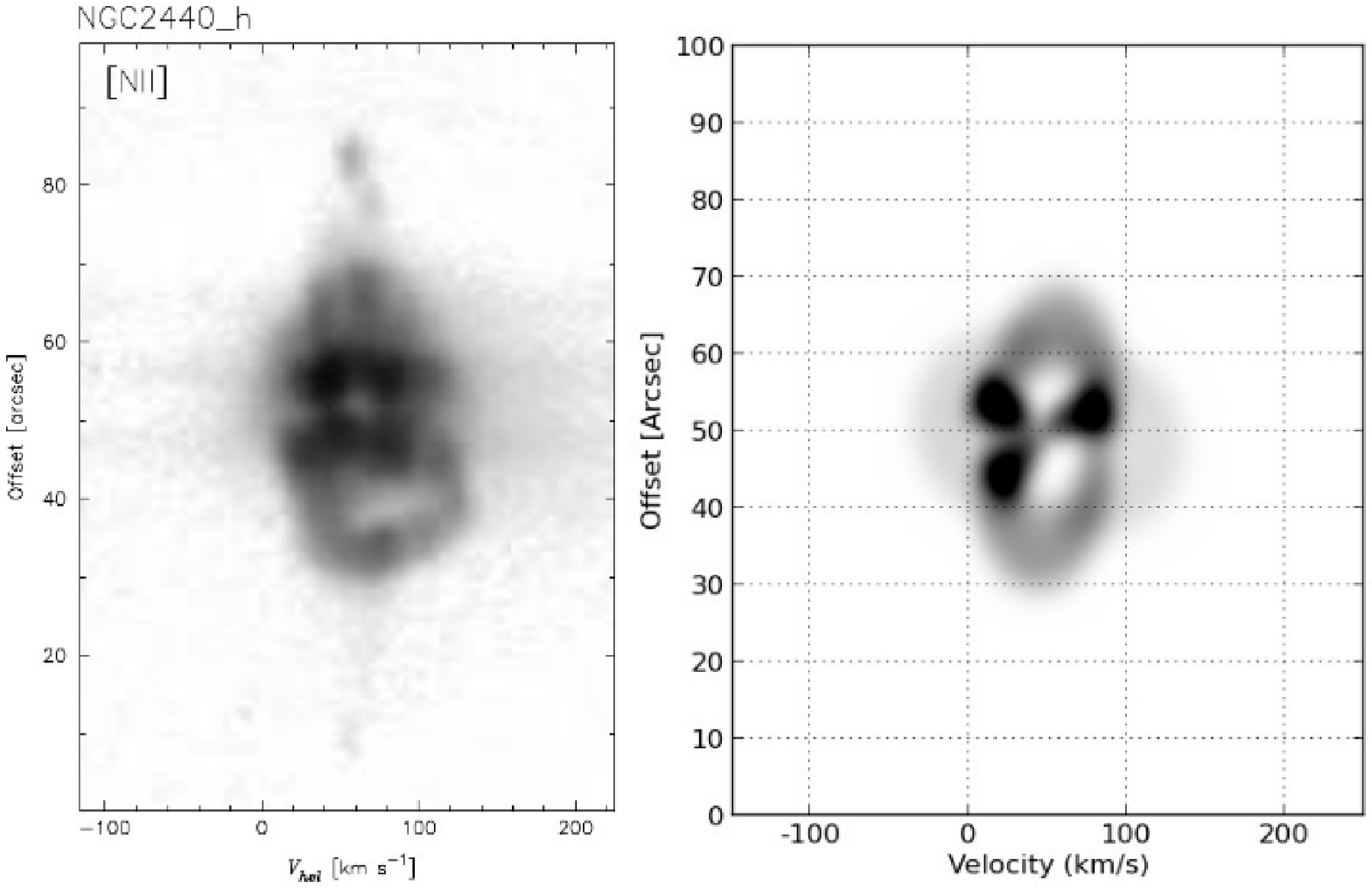}
\includegraphics[scale=0.4]{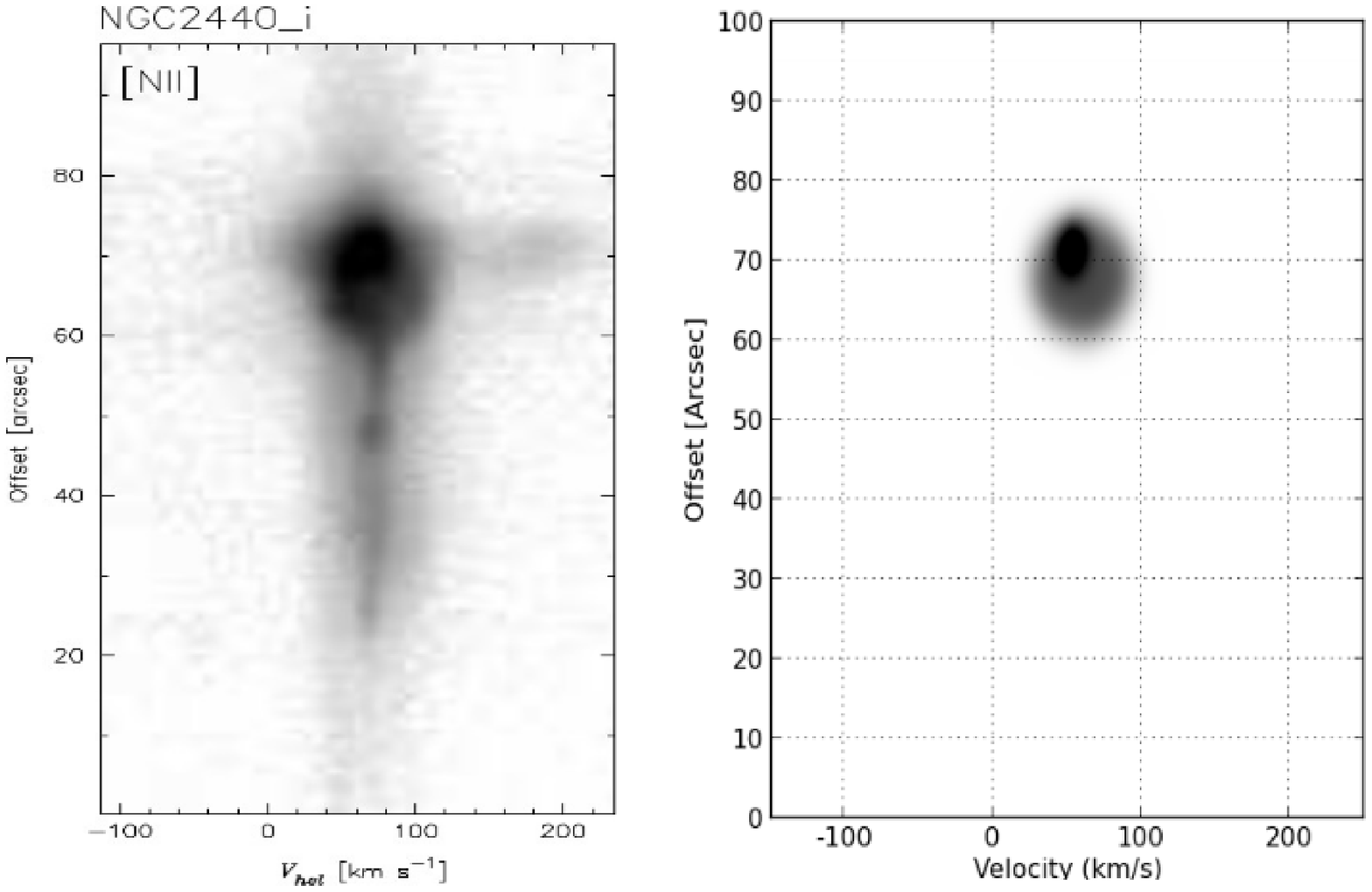}
\caption{Same of figure 3, for slit positions g, h and i.}
\end{center}
\end{figure}

\pagebreak

\begin{figure}[H]
\begin{center}
\includegraphics[scale=0.4]{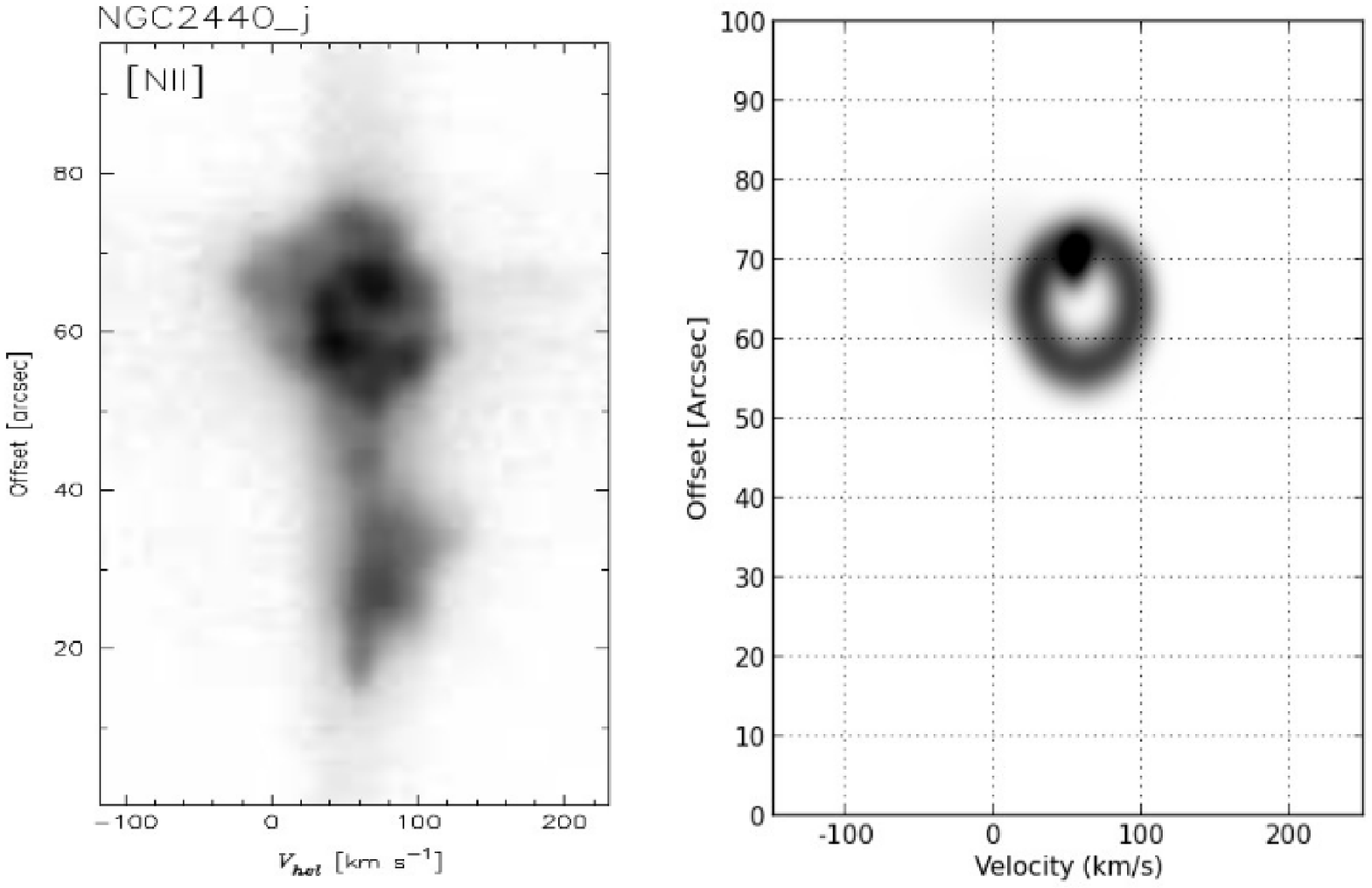}
\includegraphics[scale=0.4]{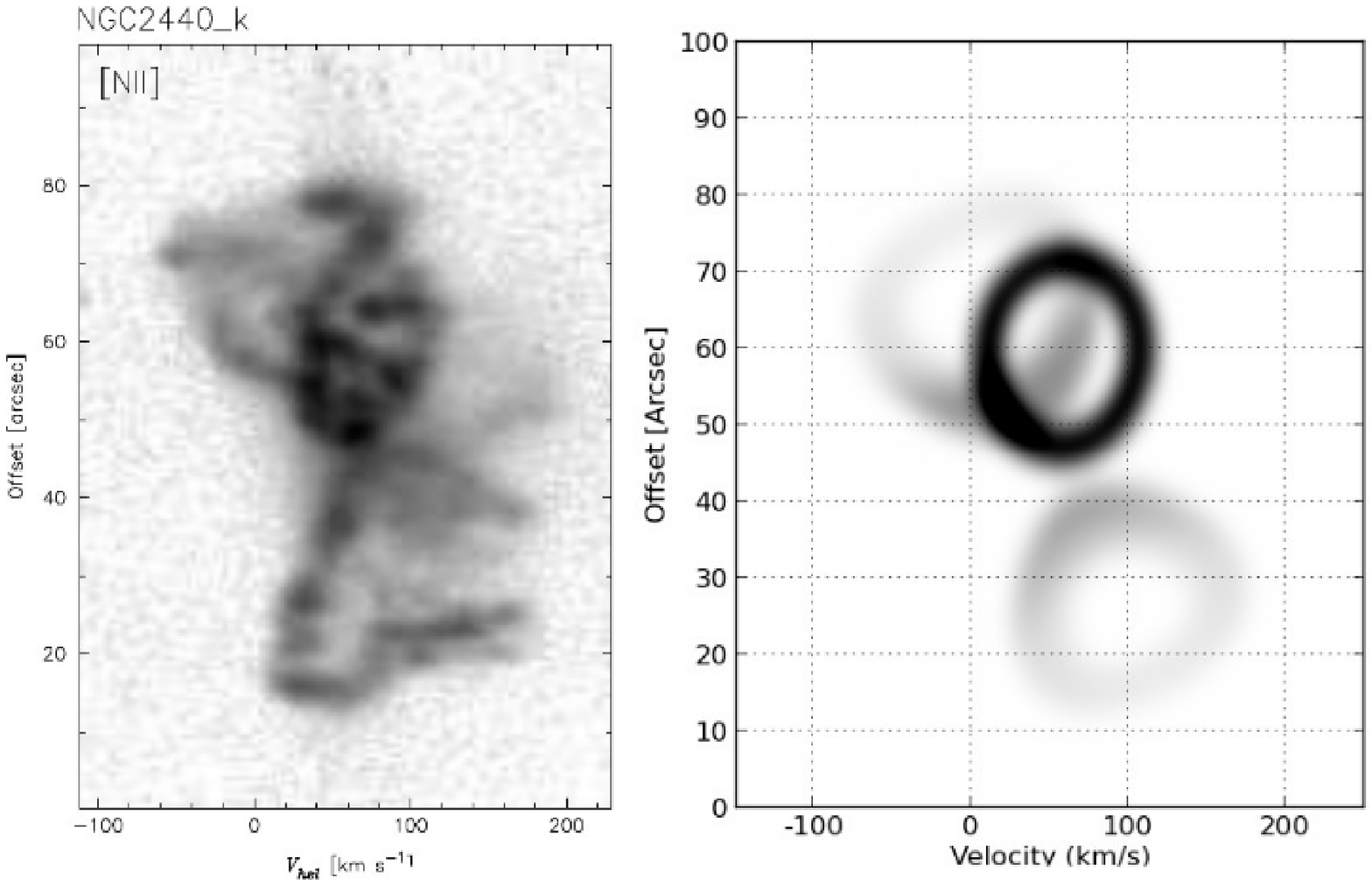}
\includegraphics[scale=0.4]{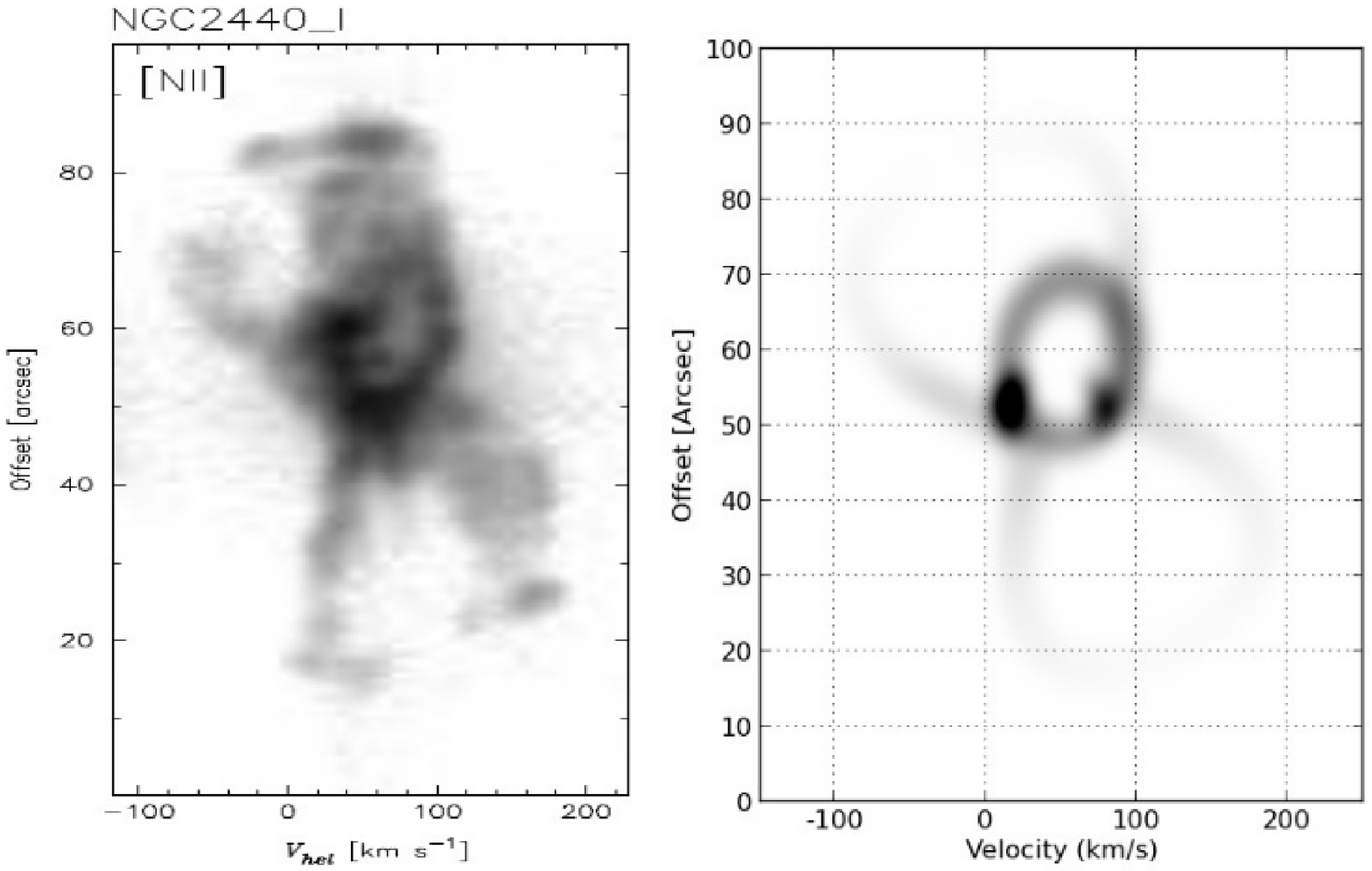}
\caption{Same of figure 3, for slit positions j, k and l.}
\end{center}
\end{figure}

\begin{figure}[H]
\begin{center}
\includegraphics[scale=0.4]{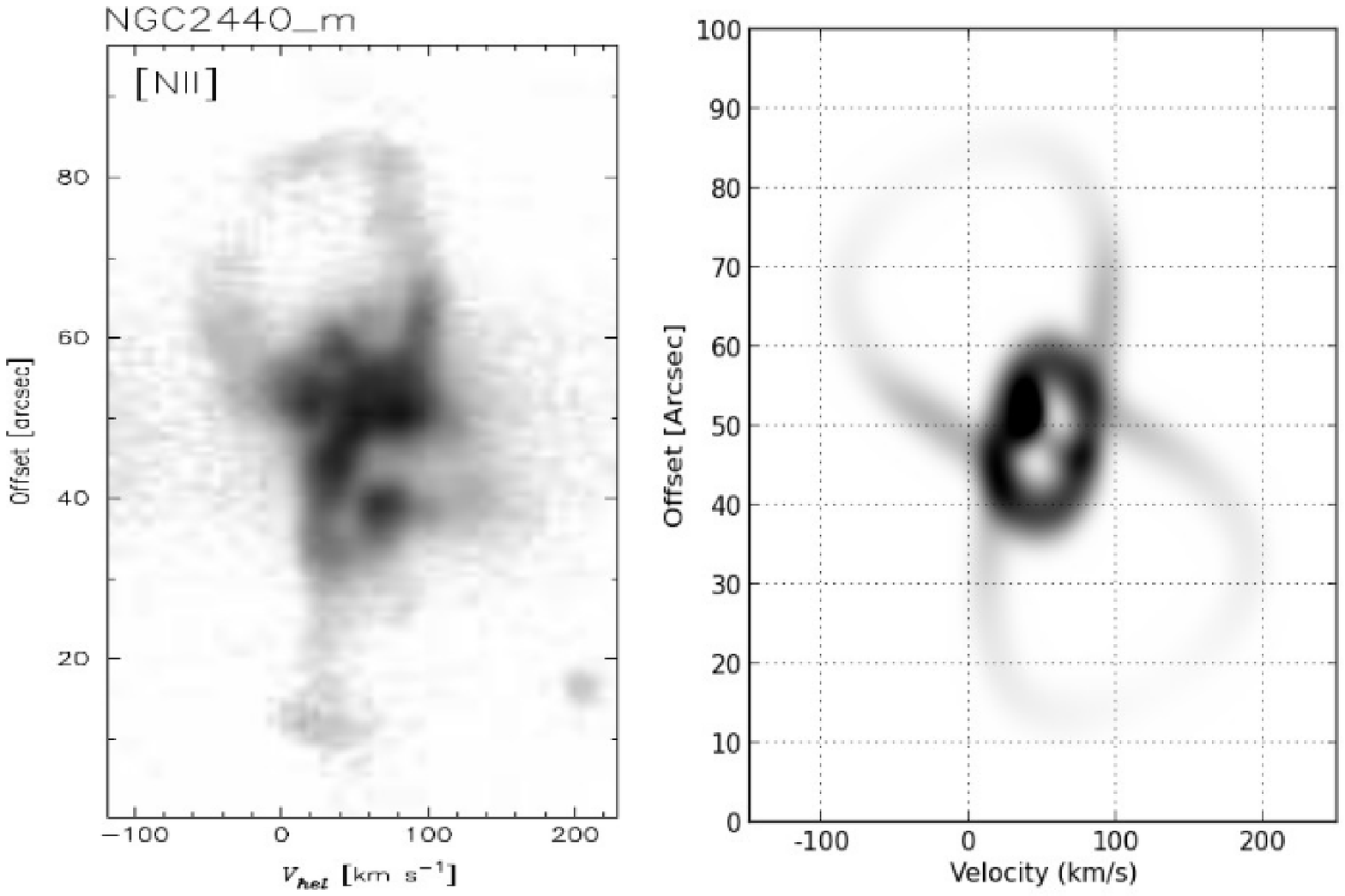}
\includegraphics[scale=0.4]{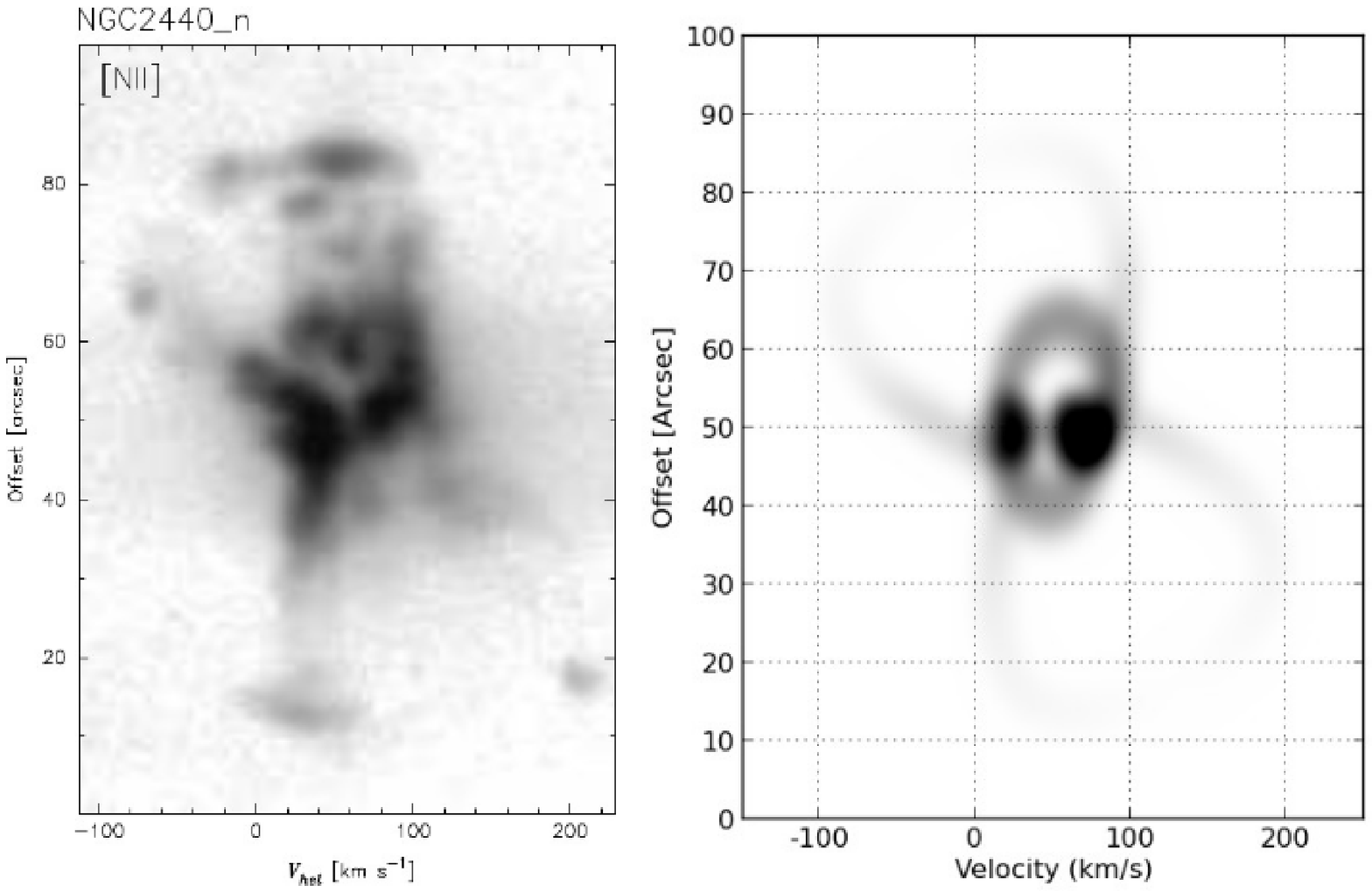}
\includegraphics[scale=0.4]{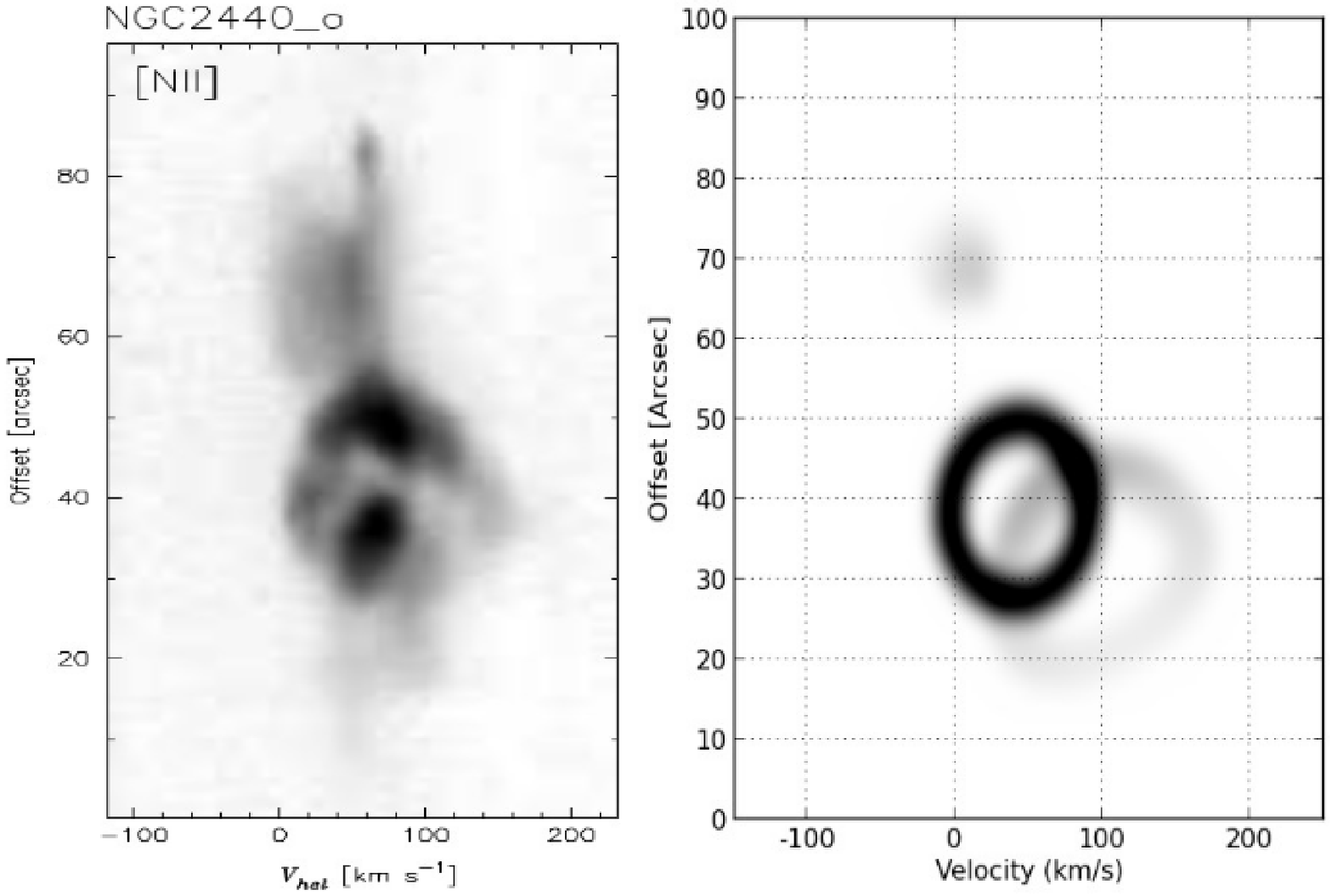}
\caption{Same of figure 3, for slit positions m, n and o.}
\end{center}
\end{figure}

\begin{figure}[H]
\begin{center}
\includegraphics[scale=0.4]{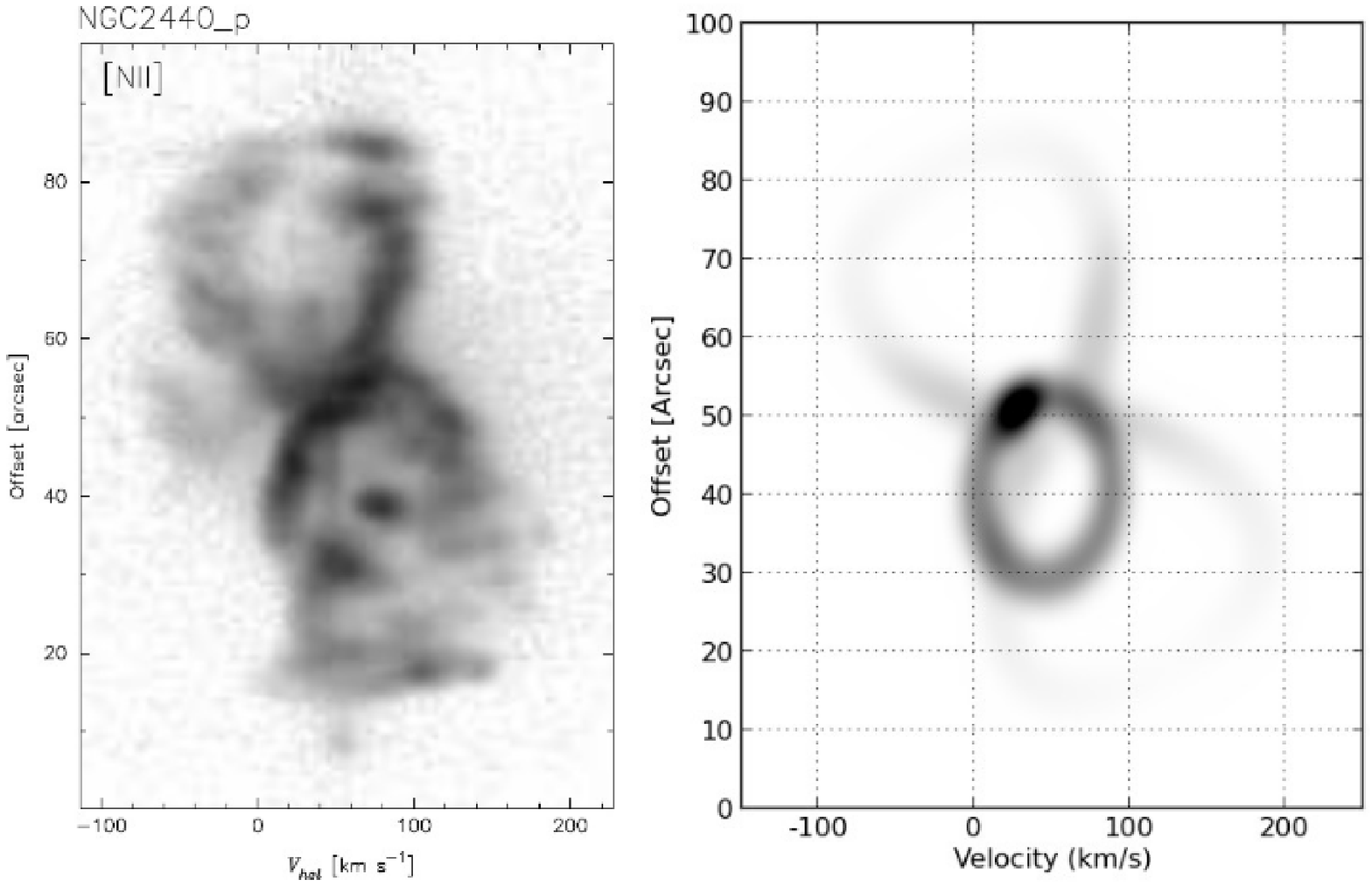}
\includegraphics[scale=0.4]{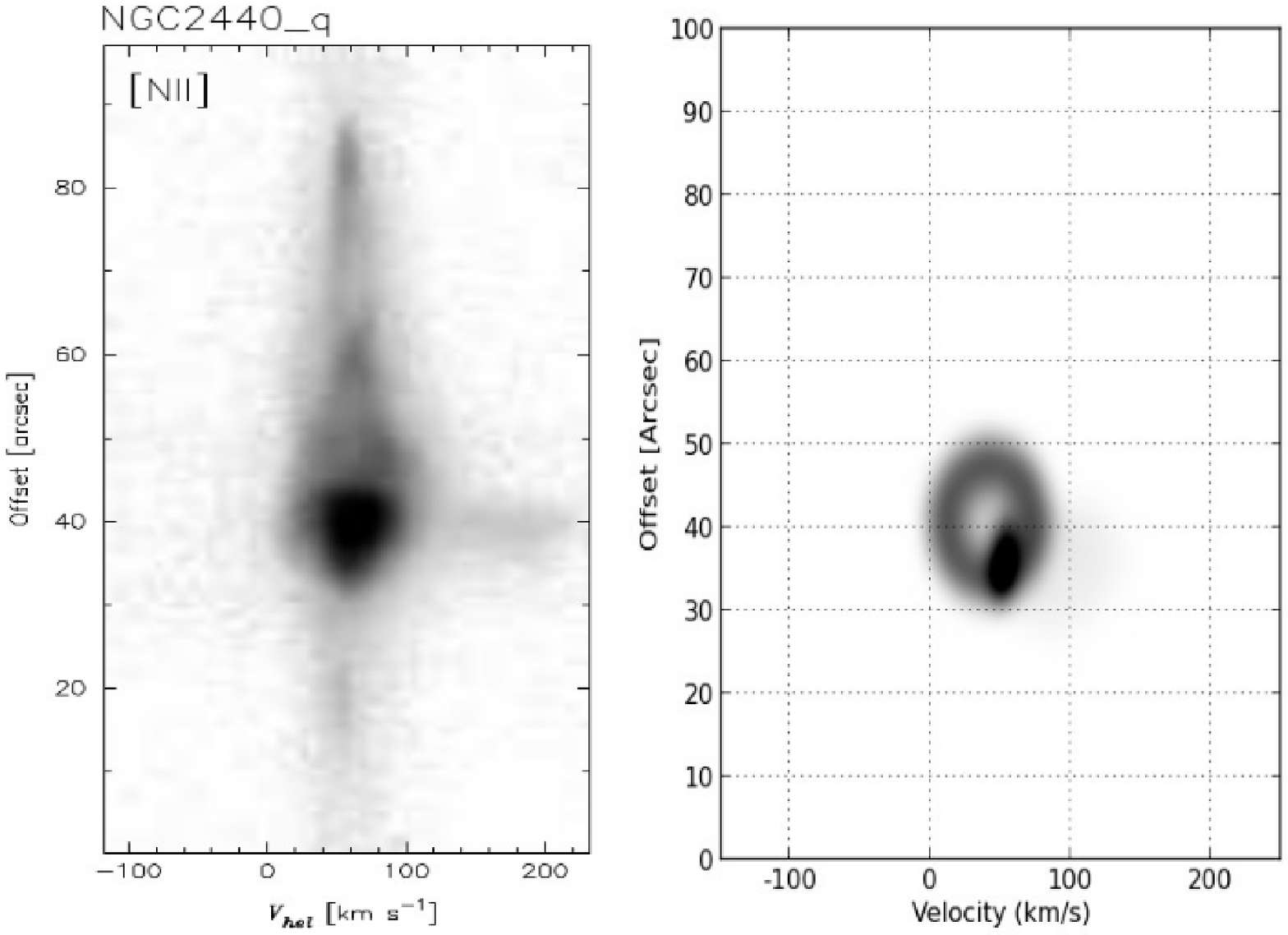}
\caption{Same of figure 3, for slit positions p and q.}
\end{center}
\end{figure}

\pagebreak

We found inclinations of the symmetry axes of the bipolar lobes with respect to the plane of the sky, respectively of $27 \pm 5$ and $-5 \pm 3$ degrees for lobes at PA 85 and PA 35 degrees.

The observational data, in particular for slit positions ``d" and ``e", suggest the presence of a toroidal structure surrounding the core of the planetary nebula. This torus is inside the inner bipolar structure at PA 35 degrees, however it is very hard to distinguish such a feature in the high angular resolution images from HST, as shown by L\'opez et al. (1998), as well in the figure 2, due to filaments and ``secondary" morphological structures. Based on the observational data and in our model, we conclude that a toroidal structure not only is there but also is strongly interacting with the environment or the radiation field. The torus is subdivided in at least three parts, as can be seen in figure 4, the P-V diagram of slit position ``d". This scenario is suggested by the diagram since the three observed substructures cannot be explained by a simple torus, what is in agreement with the results derived by \citet{2008ApJ...673..264W}.

Other small structures, such as filaments and small knots, mainly those close to the core were disregarded in the model due the lack of clear spectral signature in the kinematic profile and very complex and undefined borders in HST images.

\section{Discussion}

\subsection{Nebular Structure}

The orientation angles of the bipolar components were extracted directly from the 3D model and measured with respect to the
 plane of sky. We found inclinations of $27 \pm 5$ and $-5 \pm 3$ degrees, for bipolar components with PA of 85 and 35 degrees respectively. The presented model confirms that the bipolar lobe with PA 35 is practically in the plane of the sky. These values are different from the previous values of 40 and 0 degrees derived by L\'opez et al. (1998). This difference is probably due to the fact that previous results were derived directly from the images, and may be strongly affected by projection effects.

Looking at the big picture, excluding substructures like small knots and filaments, the model is able to reproduce the P-V diagrams for all the slit positions available in the SPM catalogue, as can be seen in figures 3 to 8. All the spectral signatures of the main components are well reproduced in the simulated P-V diagrams. Slit position ``h" shows, in P-V diagram, some substructures that are not present in our results, what is probably due to the strong segmentation in the core region, rich in filaments and other structures, which could be formed by the same processes responsible for the segmentation of the toroidal structure.

Comparing simulated and observed P-V diagrams, it is possible to see that the hypothesis of the segmented torus is in agreement with the data, however it is not possible to speculate about small segments of this torus due to the lack of observational evidence, therefore substructures of the core, in this scenario, could not be linked with the torus.

The observational data show some irregularities in the structure of the bipolar components, in particular that at PA 85 degrees, as can be seen in figure 5 for P-V diagram for slit position ``g". In several P-V diagrams  filamentary structures can be seen, however the modelled diagrams show smooth profiles, since these smaller substructures were not considered in the model.

Using the proper motion of 4.4 mas/y derived for NGC 2440 by \citet{1996PASP..108..419H}, and already used by L\'opez et al. (1998), as well as our derived velocity field for the inner structure of the nebula,  a distance of $1.8 \pm 0.5 kpc$ to the nebula was estimated, based on equation 1 from \citet{2012MNRAS.423..925A}. This value agrees with the result from \citet{1986A&A...157..171G} who derive a distance of $2.19 \pm 0.39kpc$ for this nebula.

\subsection{The origin of NGC2440 morphology}

As discussed in the introduction, there are two theoretical scenarios to explain the origin of multipolar planetary nebulae, presented by \citet{1995ApJ...455L..63L} and \citet{2013MNRAS.436..470S}. NGC 2440 shows at least two well defined bipolar components, and has another structures that are compatible with a BRET: i) the torus and the relative position of the bipolar component at PA 35 to it; ii) the possible presence of shock ionizated structures, such as two knots. Moreover, according to \citet{1997ApJS..112..487S} the central star is classified as binary, and all conditions to the formation of the nebula structure by BRET exist. Concerning the knots, low ionization lines such as [OI], [NII] and [SII] are strong indicators of shocks interaction \citep{200ApJ...0543..754C}. The high expansion velocities of these structures support this hypothesis. 

Based on previous results, mainly those after \citet{1998ApJ...493..803L}, and in the observational data combined with our models, it can be said that the BRET mechanism is still the best scenario to explain the origin of multipolar nebulae such as NGC 2440. The alternative scenario proposed by Steffen et al. (2013) is excellent to explain the formation of secondary lobes, reproducing structures such as starfish nebulae or the bipolars with many secondary lobes. However, multipolar nebulae with some or high degree of symmetry, are out of aim of their work. Therefore, no other acceptable alternative has been proposed to explain the morphology of objects such as NGC 2440.

\section{Conclusions}

This work presents the first 3D model generated in a 3D environment for NGC 2440. According to our model, the orientation angles of the bipolar lobes of the nebula with respect to the line of sign are $27\pm5$ and $-5\pm3$ degrees, respectively for the bipolar components with PA of 85 and 35 degrees. A distance of $1.8 \pm 0.5 kpc$ was derived to the nebula.  

From our models and the observational data we conclude that a segmented toroidal structure is a good description for the inner part of NGC 2440. The model based upon this hypothesis provides a good reproduction of the observed P-V diagrams, therefore it can be said that a torus, segmented at least in three parts, is in good agreement with the observational data. This result agrees with that of \citet{2008ApJ...673..264W}, who detect a molecular outflow in NGC 2440, aligned with the optical bipolar lobes.

The existence of a third bipolar lobe at PA 65 degrees, as suggested by L\'opez et al (1998) was not confirmed by our model, that was able to reproduce the global structure of the nebulae in a 3D simulation without invoking this additional component in order to reproduce the kinematical profile. However, we reinforce their suggestion that the BRET mechanism is the best alternative to explain the morphological origin of NGC 2440 due its level of symmetry.

\section{Acknowledgements}

We wish to thank S. Akras, N. Clyne and D. Gon\c{c}alves for many fruitful discussions. We wish to thank also the anonymous referee whose comments helped us to considerably increase the quality of the work. We also thank FAPESP for the financial support (Process 2010/18835-3).  P.J.A. Lago acknowledges CAPES and CNPq (Process 140803/ 2014-9) for his graduate scholarships.  

\bibliography{bibliografia.bib}

\end{document}